\documentclass[aps,prl,amsmath,amssymb,reprint,superscriptaddress]{revtex4-2}
\usepackage{graphicx}% Include figure files
\usepackage{bm}% bold math
\usepackage{threeparttable}

\begin{document}

\title{Lithium Niobate Vertical Cavity Electro-Optic Modulator}

\author{Jikun Liu$^\dagger$}
\affiliation{The Key Laboratory of Weak-Light Nonlinear Photonics, Ministry of Education, School of Physics and TEDA Applied Physics Institute, Nankai University, Tianjin 300071, People’s Republic of China}

\author{Weiye Liu$^\dagger$}
\affiliation{The Key Laboratory of Weak-Light Nonlinear Photonics, Ministry of Education, School of Physics and TEDA Applied Physics Institute, Nankai University, Tianjin 300071, People’s Republic of China}

\author{Wei Wu}
\affiliation{The Key Laboratory of Weak-Light Nonlinear Photonics, Ministry of Education, School of Physics and TEDA Applied Physics Institute, Nankai University, Tianjin 300071, People’s Republic of China}

\author{Ziang Guo}
\affiliation{The Key Laboratory of Weak-Light Nonlinear Photonics, Ministry of Education, School of Physics and TEDA Applied Physics Institute, Nankai University, Tianjin 300071, People’s Republic of China}

\author{Changrui Zhu}
\affiliation{The Key Laboratory of Weak-Light Nonlinear Photonics, Ministry of Education, School of Physics and TEDA Applied Physics Institute, Nankai University, Tianjin 300071, People’s Republic of China}

\author{Lun Qu}
\affiliation{The Key Laboratory of Weak-Light Nonlinear Photonics, Ministry of Education, School of Physics and TEDA Applied Physics Institute, Nankai University, Tianjin 300071, People’s Republic of China}

\author{Pengfei Zhu}
\affiliation{The Key Laboratory of Weak-Light Nonlinear Photonics, Ministry of Education, School of Physics and TEDA Applied Physics Institute, Nankai University, Tianjin 300071, People’s Republic of China}

\author{Yiting Zhang}
\affiliation{The Key Laboratory of Weak-Light Nonlinear Photonics, Ministry of Education, School of Physics and TEDA Applied Physics Institute, Nankai University, Tianjin 300071, People’s Republic of China}

\author{Zhihao Chen}
\affiliation{The Key Laboratory of Weak-Light Nonlinear Photonics, Ministry of Education, School of Physics and TEDA Applied Physics Institute, Nankai University, Tianjin 300071, People’s Republic of China}

\author{Qinglian Li}
\affiliation{The Key Laboratory of Weak-Light Nonlinear Photonics, Ministry of Education, School of Physics and TEDA Applied Physics Institute, Nankai University, Tianjin 300071, People’s Republic of China}

\author{Dahuai Zheng}
\affiliation{The Key Laboratory of Weak-Light Nonlinear Photonics, Ministry of Education, School of Physics and TEDA Applied Physics Institute, Nankai University, Tianjin 300071, People’s Republic of China}

\author{Hongde Liu}
\affiliation{The Key Laboratory of Weak-Light Nonlinear Photonics, Ministry of Education, School of Physics and TEDA Applied Physics Institute, Nankai University, Tianjin 300071, People’s Republic of China}

\author{Shaowei Wang}
\email{swwang@lps.ecnu.edu.cn}
\affiliation{State Key Laboratory of Precision Spectroscopy East China Normal University Shanghai 200062,China}

\author{Wei Cai}
\affiliation{The Key Laboratory of Weak-Light Nonlinear Photonics, Ministry of Education, School of Physics and TEDA Applied Physics Institute, Nankai University, Tianjin 300071, People’s Republic of China}

\author{Mengxin Ren}
\email{ren\_mengxin@nankai.edu.cn}
\affiliation{The Key Laboratory of Weak-Light Nonlinear Photonics, Ministry of Education, School of Physics and TEDA Applied Physics Institute, Nankai University, Tianjin 300071, People’s Republic of China}
\affiliation{Academy for Advanced Interdisciplinary Studies, Nankai University, Tianjin 300071, People’s Republic of China}
\affiliation{Collaborative Innovation Center of Extreme Optics, Shanxi University, Taiyuan, Shanxi 030006, People’s Republic of China}

\author{Jingjun Xu}
\email{jjxu@nankai.edu.cn}
\affiliation{The Key Laboratory of Weak-Light Nonlinear Photonics, Ministry of Education, School of Physics and TEDA Applied Physics Institute, Nankai University, Tianjin 300071, People’s Republic of China}

\begin{abstract}

Electro-optic modulators (EOMs) are vital for optical imaging and information processing, with free-space devices enabling LiDAR and beam control. Lithium niobate (LN), powered by the strong Pockels effect and scalable LN-on-insulator (LNOI) platform, has become a leading material for high-performance EOMs. Here we realize a vertical-cavity EOM in which an LN membrane is sandwiched between two photonic crystal (PhC) mirrors with integrated electrodes. The cavity supports sharp defect-mode resonances that shift efficiently under the Pockels effect, enabling strong modulation of transmission. Experiments show a depth of 43\% at ±50 V and a bandwidth of 5 MHz. This architecture combines free-space compatibility with fabrication simplicity, opening new routes to compact electro-optic platforms for ranging, holography, and beam steering.

\end{abstract}

\maketitle

Electro-optic modulators (EOMs) are indispensable in modern photonic systems, providing core functionalities for high-speed optical communications\cite{fan2020advancing, portnoi2021bandwidth}, ultrafast optical computing\cite{solli2015analog, zhou2023ultrafast, wu2025intelligent}, precision ranging\cite{lukashchuk2023chaotic, chang2024dispersive}, computational imaging\cite{lin2025high, fan2024integral, dainese2024shape, xu2024quasicrystal, liu2024broadband, fu20252nd} and emerging quantum technologies\cite{ma2025quantum, wang2020integrated, solntsev2021metasurfaces, xu2016robust}. Among the underlying mechanisms such as plasma dispersion\cite{forouzmand2019electro, park2021all}, and electro-absorption\cite{liu2008waveguide, feng201130ghz}, the Pockels effect stands out for its sub-nanosecond response times, pure refractive index modulation, and negligible optical loss\cite{abel2019large, chmielak2011pockels}. Lithium niobate (LN), in particular, combines large electro-optic coefficients, wide transparency (0.35–4.5 $\mu$m), and robust stability, making it one of the most attractive materials for EOM development\cite{weis1985lithium, boes2023lithium, gao2021long}. The recent advent of LN-on-insulator (LNOI) technology has further revitalized LN photonics, enabling compact, high-performance modulators based on Mach–Zehnder interferometers, micro-rings, micro-disks, and photonic crystals\cite{renaud2023sub, wang2022metasurface, fedotova2022lithium, wang2021efficient, sun2022self, ma2021nonlinear, qu2022giant, qu2023bright, jinfast, he2024electro, fedotova2020second, poberaj2012lithium, jia2021ion}.

Despite these advances, most LNOI modulators remain in-plane devices optimized for guided-wave operation, whose footprint and coupling requirements limit their use in free-space optical systems such as LiDAR\cite{Wang2023OnchipIO, shang2022electro, jinnonlinear, li2022progress} and computational imaging\cite{xiong2021augmented, barbastathis2019use, sinha2017lensless}. To address these constraints, compact free-space modulators have been actively explored. LN metasurfaces and hybrid architectures that integrate LN with plasmonic or dielectric nanostructures have demonstrated resonance-based electro-optic tunability\cite{gao2021electro, weigand2021enhanced, ju2022hybrid, ju2023electro, weiss2022tunable, damgaard2023nonlocal, damgaard2021electrical, damgaard2022electro, chen2025tunable, damgaard2025highly, sun2025atomic, dagli2025ghz, di2025efficient}. However, such approaches often demand sophisticated nanofabrication with subwavelength patterning and strict process control, which can limit scalability and device uniformity. By contrast, recent studies have revealed that even unpatterned LN thin films can support resonant optical modes\cite{liu2024lithium}, highlighting an alternative pathway where cavity-based geometries provide efficient and robust free-space modulation with reduced fabrication complexity.

Here, we introduce the first demonstration of a vertical-cavity EOM that harnesses defect-mode resonances in photonic crystal (PhC) to boost light–matter interaction. The device employs a LN membrane vertically sandwiched between two PhC mirrors, forming a compact free-space modulator with high-Q resonances. By incorporating surface electrodes aligned along the optical axis, we achieve a modulation depth of 43\% for $z$-polarized light at 787~nm under ±50 V drive, together with a 5 MHz modulation bandwidth. This approach not only advances thin-film LN modulators beyond metasurface architectures but also establishes a versatile platform for ultra-compact, free-space electro-optic devices.

\section{Results} 

\begin{figure}[htbp]
\centering
\includegraphics[width=85mm]{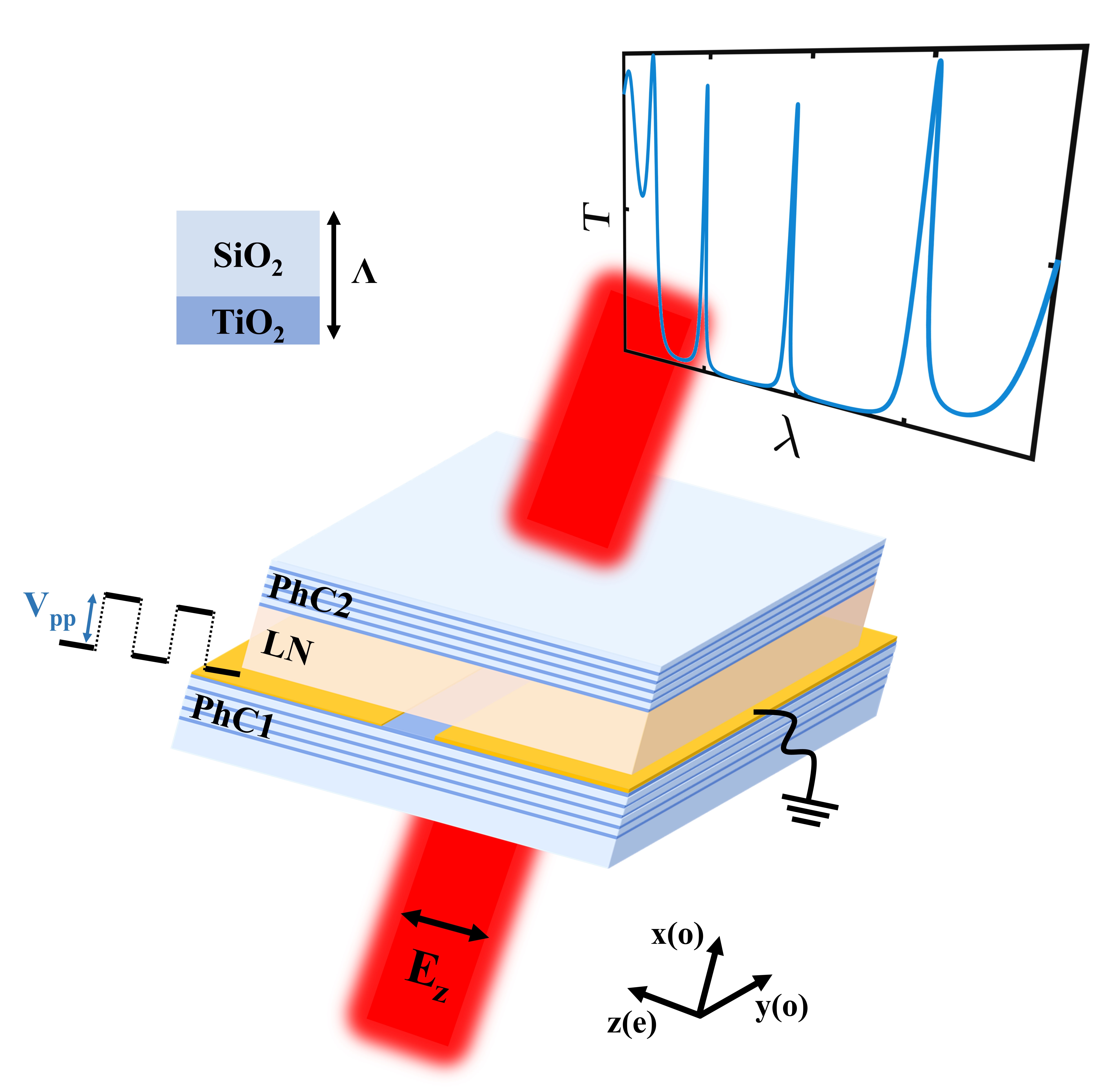}
\caption{\textbf{Schematic of the LN Vertical-Cavity EOM.} 
The device consists of a thin LN membrane vertically sandwiched between two photonic crystal (PhC) mirrors, each composed of alternating TiO$_2$ and SiO$_2$ layers. The coordinate system is aligned with the principal axes of the LN crystal, with the $z$-axis along the optic axis. Two gold electrodes are embedded at the interface between the LN membrane and the bottom PhC. This vertical-cavity configuration supports a series of resonances in the transmission spectrum. A square-wave voltage with peak-to-peak amplitude $V_{pp}$ is applied to the electrodes, while a $z$-polarized light beam impinges normally on the modulator surface, producing a modulated optical output.}
\label{Figure1}
\end{figure}

Figure 1 shows the proposed vertical cavity EOM, where an $x$-cut LN thin film is embedded between two one-dimensional photonic crystals (PhCs), forming a defect cavity that supports sharp transmission resonances. Each PhC consists of alternating titanium dioxide (TiO$_2$) and silicon dioxide (SiO$_2$) slabs, while the LN film has a thickness of 1~$\mu$m (NANOLN Co. Ltd.). Gold electrodes, with 50~nm-thickness and separated by a gap of 10~$\mu$m, are fabricated between the LN layer and the lower PhC, through which a square-wave voltage ($V_{pp}$) is applied. The resulting driving electric field is aligned with the optic axis ($z$-direction), and by matching the incident polarization to this axis, the Pockels effect is maximized. Consequently, the LN refractive index is modulated, leading to resonance shifts in the transmission spectrum.

The refractive index ellipsoid of LN under $V_{pp}$ is given by:

\begin{equation}\label{eq1} \footnotesize
\left(\frac{1}{n_o^2}+\gamma_{13}\frac{V_{pp}}{d} \right)x^2 + \left(\frac{1}{n_o^2}+\gamma_{13}\frac{V_{pp}}{d} \right)y^2 + \left(\frac{1}{n_e^2}+\gamma_{33}\frac{V_{pp}}{d} \right)z^2
= 1.
\end{equation}

\noindent where $n_o$ and $n_e$ are the ordinary and extraordinary refractive indices, $d$ is the electrode spacing, and $\gamma_{33}$ = 31.45~pm/V is the dominant electro-optic coefficient. The extraordinary index changes as \cite{thomaschewski2022pockels}:

\begin{equation}\label{eq2}
\vert\Delta n_e\vert = \frac{1}{2}n_e^3\gamma_{33}\frac{V_{pp}}{d}.
\end{equation}

\begin{figure}[htbp]
\centering
\includegraphics[width=85mm]{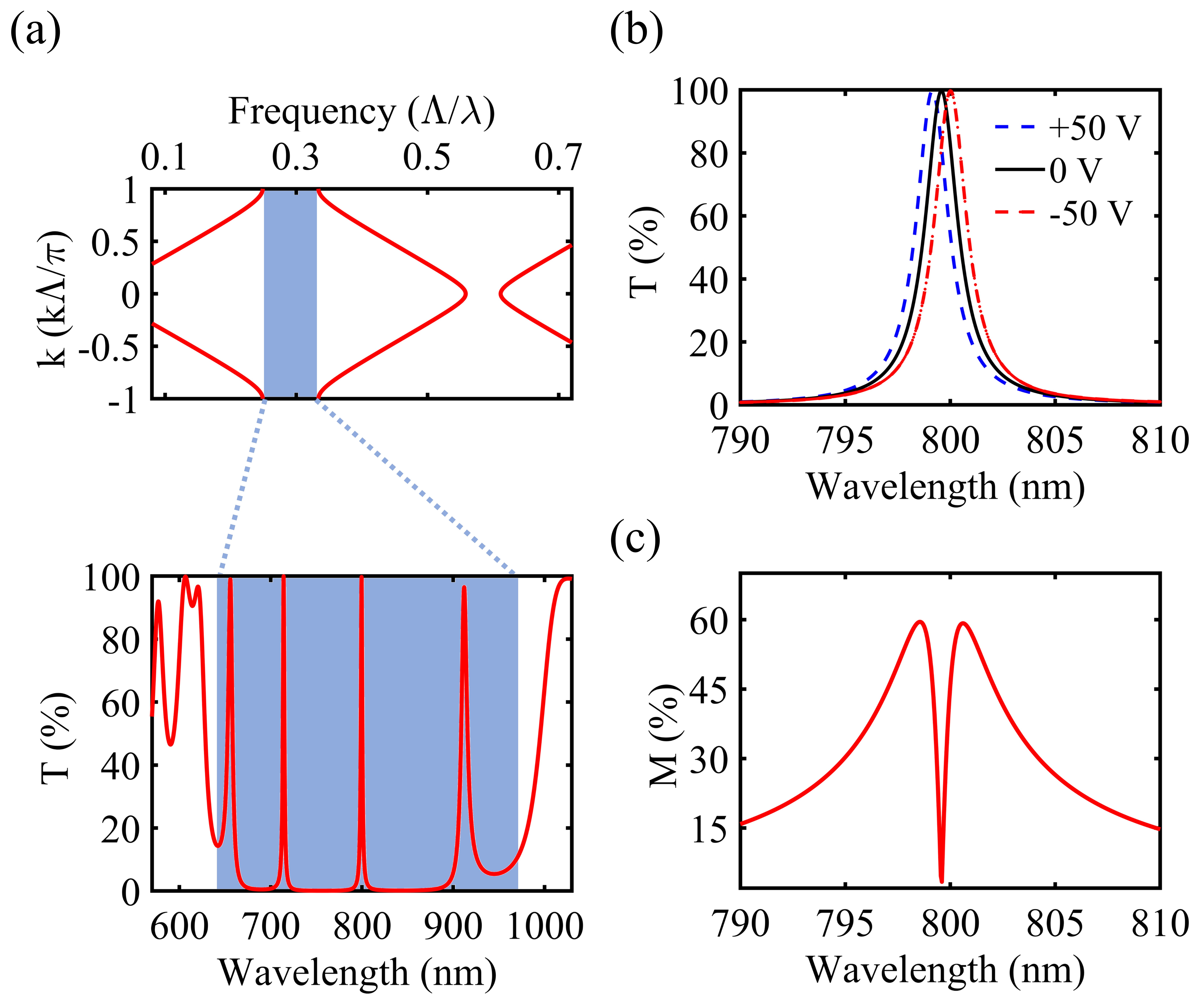}
\caption{\textbf{Band structure engineering and simulated performance of the vertical cavity EOM.}
(\textbf{a}) Top: Simulated photonic band structure of the PhC, showing a bandgap centered near a normalized frequency of 0.28. Bottom: The 1~$\mu$m LN defect introduces cavity modes around 700, 800 and 920~nm. (\textbf{b}) Transmission spectra around 800~nm under different modulation voltages: dashed blue, +50~V; solid black, 0~V; dashed red, -50~V. (\textbf{c}) Simulated modulation depth ($M$) corresponding to the $\pm$50~V voltage switching.}
\label{Figure2} 
\end{figure}

The PhC parameters were optimized by finite element method (COMSOL Multiphysics), using optical constants of SiO$_2$, TiO$_2$ and LN measured by spectroscopic ellipsometry (Accurion EP4) \cite{liu2021machine}. With $d_{\text{SiO}_2}$ = 160~nm and $d_{\text{TiO}_2}$ = 65~nm, the calculated photonic band structure exhibits a clear bandgap at normalized frequency 0.28. Inserting the 1~$\mu$m LN defect layer produces resonances around 700, 800 and 920~nm under $z$-polarized excitation (Figure 2a). We focus on 800~nm resonance that coincides with the wavelength range (770–840~nm) of the laser used in experiments. 

Figure 2b shows the simulated transmission spectra under applied voltages of +50~V (blue), 0~V (black), and –50~V (red). Because $\gamma_{33}$ is negative, a positive bias reduces $n_e$, resulting in a blue shift, whereas a negative bias induces a red shift. The corresponding modulation depth, defined as $M(\lambda)=1-\frac{T_{min}(\lambda)}{T_{max}(\lambda)}$, is shown in Figure 2c \cite{rahm2013thz}. The $M(\lambda)$ exhibits an M-shaped profile at $\pm$50~V, with peaks at 798.5~nm and 800.6~nm, coinciding with the wavelengths where slopes of transmission spectrum in Figure 2b are steepest \cite{chen2018plasmonic, argyropoulos2015enhanced}.

\begin{figure}[htbp]
\centering
\includegraphics[width=85mm]{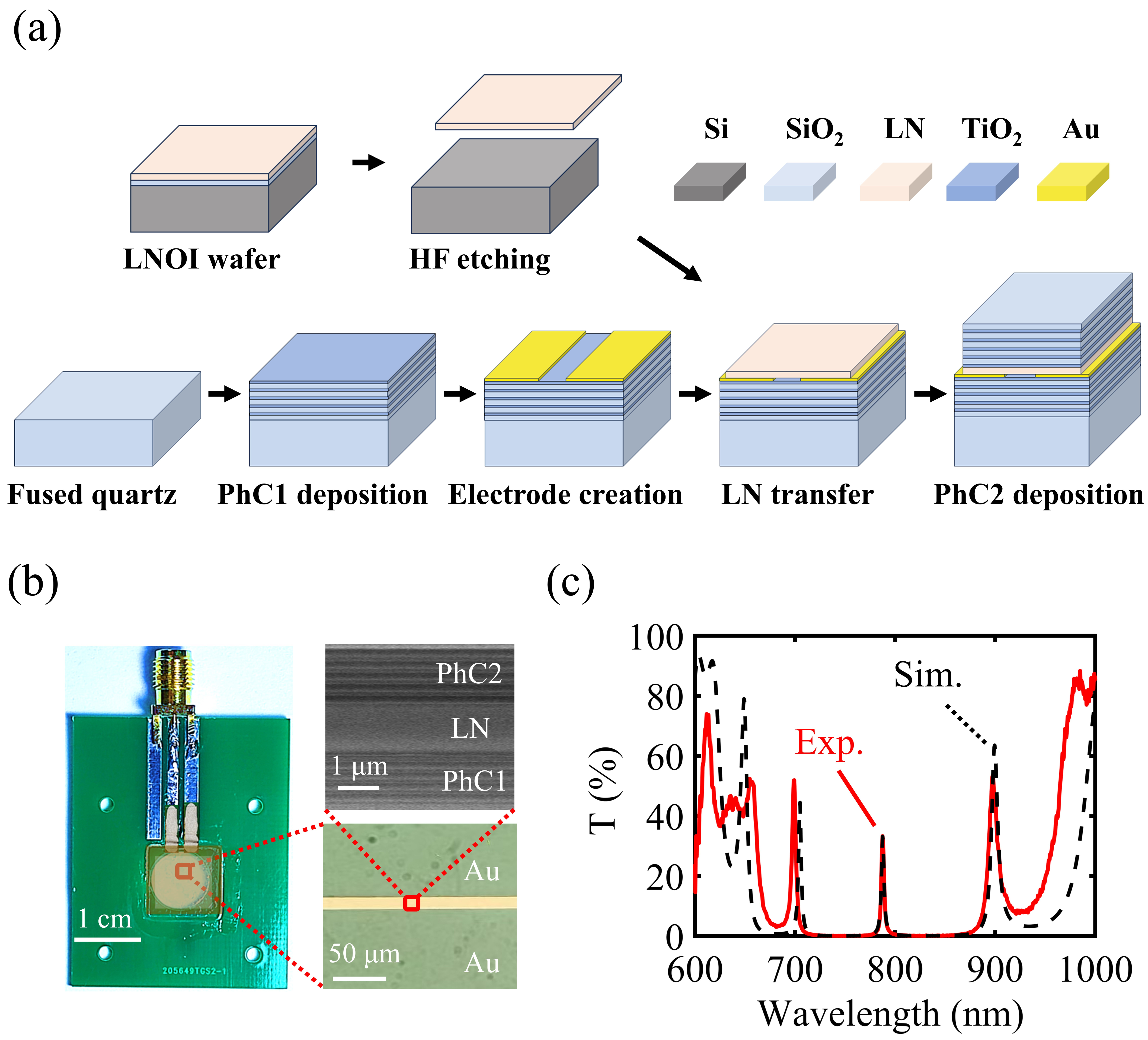}
\caption{\textbf{Fabrication of the EOM.} 
(\textbf{a}) The fabrication process: starting from an LNOI wafer with an $x$-cut LN film bonded to a silicon substrate via a SiO$_2$ buffer layer, the LN film was released using HF etching. Gold electrodes with a 10~$\mu$m pitch were patterned on a fused silica substrate pre-coated with PhC1 (alternating TiO$_2$/SiO$_2$ layers). The LN film was then transferred and aligned onto the electrodes, followed by deposition of PhC2 on the LN to form the vertical cavity. (\textbf{b}) Left: packaged vertical EOM. Bottom right: optical microscopy of the modulator surface. Top right: SEM image showing the detailed cross-section of the LN vertical cavity. (\textbf{c}) Experimental transmission spectra (red curves) under normally incident $z$-polarized light, compared with simulations (black dashed curves) incorporating fabrication-induced optical losses.}
\label{Figure3}
\end{figure}

The fabrication procedure of the sample is summarized in Figure 3a. The process started with hydrogen fluoride (HF) etching of a commercial LNOI wafer, consisting of an $x$-cut LN film bonded to a silicon substrate through a SiO$_2$ buffer\cite{bai2024wavelength}. In parallel, gold electrode pads with a 10~$\mu$m gap were patterned on a fused silica substrate already coated with the first PhC, using standard photolithography, metal deposition, and lift-off. The released LN film was then transferred onto this substrate and aligned such that its optic axis was along the line connecting the two electrodes. A second PhC layer was subsequently deposited on top of the LN film, completing the vertical cavity. The sample was then mounted onto a printed circuit board (PCB), and the electrodes were connected to the PCB pads using conductive silver adhesive. The final device was interfaced with external signals through an SMA port (Figure 3b, left panel). Optical microscopy clearly revealed the electrode and device regions (lower right panel), while SEM image provided detailed views of the structural cross-section (upper right panel).

The transmission spectrum was experimentally measured using a commercial microscopic spectrometer (IdeaOptics Co., Ltd.), as shown by the red curve in Figure 3c. A clear resonance appears near 788~nm, slightly blue-shifted from the designed value (Figure 2b), likely due to the fabrication imperfection, such as thickness variation of PhCs and LN film. Compared with the simulated spectrum in Figure 2a, the experimental resonance is broader, which we attribute to interface roughness and scattering losses. To account for these effects, an imaginary component ($\kappa$ = 0.00275) was added to the refractive index of LN in the simulation (Supporting Information, Section 1). The adjusted result, shown by the black dashed curve in Figure 3c, closely matches the experimental spectrum. 

\begin{figure}[htbp]
\centering
\includegraphics[width=85mm]{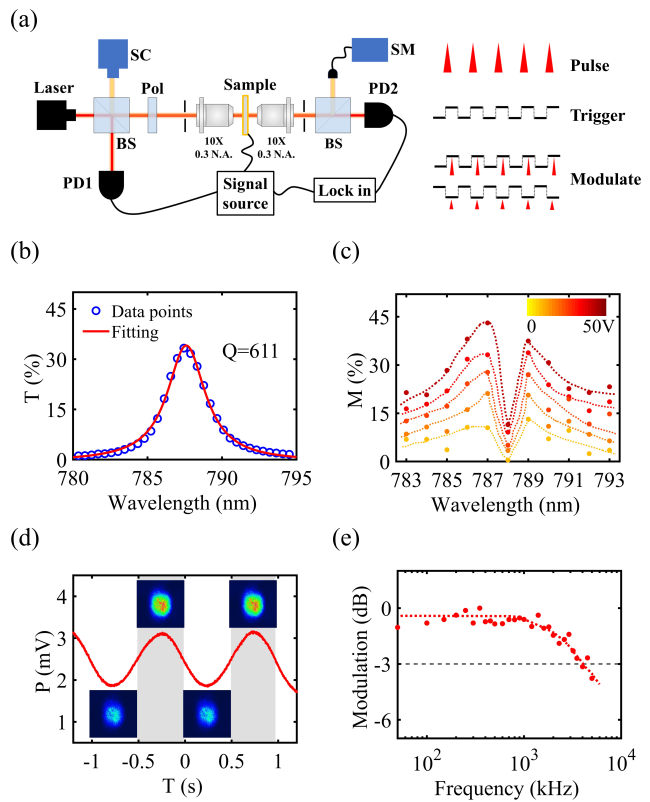}
\caption{\textbf{Experimental characterization of the vertical cavity EOM.} 
(\textbf{a}) Schematic of the optical setup. The tunable laser is directed through a polarizer (Pol) and a micro-focusing system with dual-aperture filtering to illuminate the modulator, which is simultaneously driven by a synchronous electrical signal. Spectral characteristics are achieved through the combined use of a supercontinuum source (SC), spectrometer (SM), and beam splitter (BS). A bypass photodetector (PD1) captures the pulse signal and couples with the signal source to generate a synchronous modulation drive signal. High-level and low-level pulse modulation is achieved by adjusting the phase delay. The end photodetector (PD2) collects the output signal, and the modulation results are analyzed via a phase-locked amplifier. (Illustration: Pulse in-phase operation timing diagram). (\textbf{b}) Transmission spectrum of the sample measured by the supercontinuum light source spectral testing system. Blue circles: experimental data; red curves: Fano-fitted spectra, yielding Q-values of 611. (\textbf{c}) Measured modulation depth ($M$) from 0 to ±50~V in 10~V steps. Dots: experimental data; dotted lines: visual guide. (\textbf{d}) Optical power modulation under differential frequency excitation, showing representative intensity profiles at maximum (top inset) and minimum (bottom inset) modulation states. (\textbf{e}) Modulation bandwidth of devices under optical power modulation with multiple-frequency differential excitation.}
\label{Figure4}
\end{figure}

To characterize the electro-optic modulation performance of the device, a tunable pulsed laser polarized along the $z$-axis (TUN-TiN, repetition rate: 50 kHz, linewidth: $<$ 40 pm, wavelength step: 1 nm) was employed as the light source. The laser beam was focused onto the sample using a 10× objective lens (numerical aperture, NA = 0.30) and collected at the transmission side by an identical objective. To suppress high-spatial-frequency components, apertures were placed at both Fourier planes before and after the objective pair for spatial filtering. The front-end photodetector (PD1, Thorlabs DET36A/M; see Fig. 4a) was used to record the pulsed signal and to generate a synchronized modulation signal in conjunction with the signal source (inset, middle right of Fig. 4a). The signal source consisted of an arbitrary waveform generator (Agilent 33250A) and a high-voltage amplifier (Falco WMA-300). By tuning the phase delay of the driving signal, high- and low-level pulse modulation could be achieved (lower-right inset in Fig. 4a, further details are provided in Section 2 of the supplementary materials).

The output modulation signal was collected by a second photodetector (PD2, Fig. 4a) and analyzed using a lock-in amplifier (Zurich UHFLI). In addition, to enable in-situ transmission spectral characterization, a supercontinuum light source (NKT Photonics SuperK Extreme) was coupled into the system via a beam splitter and analyzed by a fiber-coupled spectrometer (IdeaOptics). The measured transmission spectrum is shown as blue dots in Fig. 4b. A Fano-resonance fit (red curve) yields a quality factor (Q) of 611, corresponding to a full width at half maximum (FWHM) of approximately 1.3 nm.

EO modulation was then investigated under synchronized optical–electrical operation. Single-wavelength modulation spectra were recorded using the pulsed laser source under applied voltages from 0 to ±50 V in 10 V increments. The measured $M(\lambda)$ curves (color-coded from light to dark) exhibited the expected M-shaped profile with peaks at 787 nm and 789 nm (Figure 4c), in good agreement with simulations in Figure 2d. The maximum modulation depth reached 43\% at 787 nm under ±50 V.Time-domain measurements confirm the modulation characteristics of the device at 50 kHz (Fig. 4d). The actual repetition frequency of the laser is 49.999 kHz, resulting in a slight frequency mismatch of approximately 1 Hz with the externally applied 50 kHz modulation signal. This leads to a slow envelope modulation with a period of about 1 s, which is consistent with the externally imposed waveform. The intensity modulation was visualized using a CCD (Beamage-4M), and the insets display the contrast between the maximum and minimum states (see Supplementary Video).

Furthermore, delayed sampling was employed to apply modulation at integer multiples of 50 kHz to characterize the high-frequency response of the device. The time-domain optical intensity distributions at various modulation frequencies were recorded (see Supplementary Section 3), and the corresponding frequency response of the modulation signal is shown in Fig. 4e. Owing to the bandwidth limitation of the high-voltage amplifier, the measured modulation intensity response exhibits a 3 dB bandwidth of approximately 5 MHz, which is considerably lower than the intrinsic bandwidth of the device. Vector network analysis presented in the Supplementary Section 4 indicates that the intrinsic bandwidth of the device extends into the GHz regime.

In addition, the results were compared with previously reported LNOI-based free-space modulators (see Supplementary Section 3). Despite the structural simplicity of our device, its overall performance remains highly competitive among similar works. With further improvements in fabrication precision and measurement techniques, the device is expected to demonstrate even greater potential.

\section{Discussion}

In summary, we demonstrate the LN vertical cavity EOM, where a thin LN layer is vertically confined between two PhC slabs to form defect-mode resonances that strongly enhance light–matter interaction. The carefully engineered structure achieves a Q-factor above 600, and generates a resonance linewidth as narrow as ~1.3 nm. Exploiting the electro-optic effect of LN, the device delivers a modulation depth of 43\% for $z$-polarized light at 787 nm under a ±50 V alternating voltage, enabled by precise resonance tuning. This planar free-space compatible architecture further highlights the potential of this platform for ultra-compact, high-performance electro-optic components in beam steering, dynamic holography, and spatial light modulation.

\medskip
\noindent\textbf{Data availability}\par
All relevant data supporting the results of this study are available within the article and its supplementary information files. Further data are available from the corresponding authors upon request.

\medskip
\noindent \textbf{Acknowledgements} \par
This work was supported by National Key R\&D Program of China (2022YFA1404800, 2023YFA1407200); National Natural Science Foundation of China (12222408, 12304423, 12304424); Natural Science Foundation of Tianjin (24JCJQJC00070); China Postdoctoral Science Foundation (2022M721719, 2022M711710); Guangdong Major Project of Basic and Applied Basic Research (2020B0301030009); 111 Project (B23045); PCSIRT (IRT0149); Fundamental Research Funds for the Central Universities. We thank Nanofabrication Platform of Nankai University for fabricating samples.

\medskip
\noindent \textbf{Competing interests} \par
The authors declare no conflict of interest.

\bibliography{Manuscript.bib}
\bibliographystyle{style}

\end{document}